  \documentclass[prl,aps,twocolumn,showpacs]{revtex4}
\usepackage[dvips]{graphicx}
\usepackage{amsmath,  amssymb, array, graphics,  amsfonts, eucal}
\usepackage{dcolumn}
\begin{document}
\newcommand{\mc}[1]{\mathcal{#1}}
\newcommand{\E}{\mc{E}}
\topmargin=-15mm

\title {\bf
Longitudinal dielectric permeability of the quantum
degenerate collisional plasmas}

\author{\bf Anatoly V. Latyshev and Alexander Yushkanov}

\affiliation{Department of Mathematical Analysis and
Department of Theoretical Physics, Moscow State Regional
University,  105005, Moscow, Radio st., 10--A}

\begin{abstract}
    Dielectric permeability of the degenerate electronic gas for
the collisional plasmas is found.
    The kinetic equation of  Wigner --- Vlasov --- Boltzmann
with integral of collisions in relaxation form
in coordinate space is used.
    We will notice that dielectric permeability with using of the
relaxation equation in the
momentum  space  has been received by Mermin.

\noindent {\bf Keywords: Degenerate Electron Gas, Dielectric
Permeability and
Conduc\-ti\-vi\-ty, Collision Integral, Lindhard Function, Kohn's
Singularities.}
   \end{abstract}
\pacs{50, 52.25.Dg Plasma kinetic equations, 52.25.-b Plasma properties}
\date{\today}
\maketitle

\section{I. Introduction}

In the present work formulas  for
conductivity and for dielectric permeability of quantum
electronic plasma are deduced.

Dielectric permeability is one of the major
plasma characteristics. This quantity is necessary for
description of process of propagation and attenuation of the plasma
oscillations, skin effect, the mechanism of
electromagne\-tic waves penetration
in plasma \cite{Shukla1} -- \cite {Silin-Ruh}, and for
analysis of other problems in plasma physics.

Dielectric permeability in the collisionless quantum
gaseous plasma was studied by many authors (see, for example,
\cite {Klim}-\cite {Arnold}). In work \cite {Manf}, where
the one-dimensional case of the quantum plasma is investigated,
importance of derivation of dielectric permeability with
use of the quantum kinetic equation with integral
collisions in the form of BGK - models (Bhatnagar, Gross, Krook)
\cite {BGK} was marked.
The present work is devoted to performance of this  problem.

In the present work for a derivation of dielectric permeability
quantum kinetic Wigner --- Vlasov --- Boltzmann equation
(WVB--equation) with  collision integral
in the form of $ \tau $--models is applied.
Such collision integral is named BGK--collision integral.

The WVB--equation is written for   Wigner function,
which is analo\-gue of distribution function of
electrons for quantum plasma
(see \cite {Wigner} and \cite {Hillery}).

The most widespread method of investigation of quantum plasmas is the
method of Hartree --- Fock  or a method equivalent to it,
namely, the method of Random Phase Approximation
\cite {Lif}, \cite {Plaz}.
In work \cite {Lind} this method  has been applied to receive expression for
dielectric permeability of quantum plasma in $ \tau $--approach.
However, in work \cite {Kliewer} it is shown, that  expression
received in \cite {Lind} is noncorrect,
as does not turn into classical expression under a condition, when
quantum amendments can be neglected. Thus in work \cite {Kliewer}
empirically corrected expres\-sion  for
dielectric permeability of quantum plasma, free from
the specified lack has been offered.
By means of this expression authors investigated
quantum amendments to optical properties of metal
\cite {Kliewer2}, \cite {Kliewer3}.

Dielectric permeability of quantum plasma  is widely
used also for studying the screening of the electric field and
Friedel oscillations (see, for example, \cite {Kohn1} - \cite {Harr}).
In the work \cite {Eminov}  screening of the Coulomb fields
in magnetised electronic gas has been is studied.

In theory of quantum plasma there exist two essentially
various possibilities of construction of the relaxation kinetic
equation in $ \tau $ -- approximation: in the space of impulses (in the
space of Fourier images of the distribution function) and in the
space of coordinates. On the basis of the relaxation kinetic
equations in the space of momentum Mermin \cite {Mermin}
has carried out consistent derivation of the dielectric
permeability for quantum collisional plasma in 1970 for the first time.

In the present work expression for the longitudinal
dielectric permeability
 with use of the relaxation equations in space of coordinates is deduced.
If in the received expression we make
Planck constant converge  to zero ($ \hbar\to 0$), we will receive
exactly classical expression of dielectric permeability of
degenerate plasma. Various limiting cases of the
dielectric permeability are investigated.
Comparison with Mermin's result is carried out also.

\section{II. Solution of the kinetic equation}

We consider the kinetic Wigner --- Vlasov --- Boltzmann
equation \cite{Gurov}:
$$
\dfrac{\partial f}{\partial t}+\mathbf{v}\dfrac{\partial f}{\partial
\mathbf{r}}=-\dfrac{ie}{\hbar}W[U,f]+B[f,f].
$$

Here $e$ is the charge of electron, $W[U,f]$ is the functional
of Wigner --- Vlasov for the scalar potential $U$,
$$
W[U,f]=\dfrac{1}{(2\pi)^3}\int \Big[U\Big(\mathbf{r}+\dfrac{\hbar
\mathbf{b}}{2},t\Big)-
U\Big(\mathbf{r}-\dfrac{\hbar \mathbf{b}}{2},t\Big)\Big]
\times
$$
$$
\times
f(\mathbf{r},\mathbf{p'},t)\exp(i\mathbf{b}(\mathbf{p'}-\mathbf{p}))\,d^3
b\, d^3p',
\eqno{(1.1)}
$$
$\mathbf{b}=\{b_x,b_y,b_z\}$ is the vector,
$f=f(\mathbf{r},\mathbf{p},t)$ is the Wigner function for
electrons, $\hbar$ is the Planck constant,
 $B[f,f]$ is the  collision integral.

Collision integral for  quantum plasma in general case can
have rather complex form.
In particular, it can be non-local by coordinates as well.
A limiting case of such quantum
non--locality is considered in \cite {Mermin}. In the present work
the case when it is possible to present collision integral
in a local form is considered.
Particularly, we will consider collision integral  representation
in a form of standart model BGK--collision integral
(Bhatnagar --- Gross --- Krook) \cite {Krook}.
Then the previous equation will be written in the following form:
$$
\dfrac{\partial f}{\partial t}+\mathbf{v}\dfrac{\partial f}{\partial
\mathbf{r}}=-\dfrac{ie}{\hbar}W[U,f]+
$$
$$+\nu \Big[f_{eq}(\mathbf{r},\mathbf{p},t)-f(\mathbf{r},\mathbf{p},t)\Big].
\eqno{(1.2)}
$$

This equation describes behaviour of the collisional degenerate quantum
plasma.

Here $\nu$ is the effective
scattering frequency  of electrons (in particular, on impurities),
$f_{eq}$ is the equilibrium Fermi --- Dirac distribution function
of electrons. Further we will consider  the case of
degenerate quantum
plasma.
 Then the equilibrium
distribution function can be expressed in terms of Heaviside function
$$
f_{eq}=\Theta(\E_{eq}(\mathbf{r},t)-\E),
$$
the function  $\Theta(x)$ is the function of Heaviside,
$$
\Theta(x)=\left\{\begin{array}{c}
              1,\quad x>0, \\
              0, \quad x<0,
            \end{array}\right.
$$
$\E$ is the kinetic energy of electrons,
$\E=\dfrac{mv^2}{2}=\dfrac{p^2}{2m}$,
$$
\E_F(\mathbf{r},t)=\dfrac{mv_F^2(\mathbf{r},t)}{2}=
\dfrac{p_F^2(\mathbf{r},t)}{2m}
$$
is the perturbed Fermi energy of the electrons, $\mathbf{p}=m\mathbf{v}$
is the momentum of the electron, $\mathbf{p}_F$  is the momentum
of the electron on the Fermi surface. We assume that Fermi
surface is spherical.

Let's consider, that distribution electron function depends on one
spatial coordinate $x$, time $t$ and momentum $\mathbf {p}$,
and the electric potential depends on one spatial coordinate
$x$ and time $t$.
Then the equations (1.1) and (1.2) can be written in a form:
$$
\dfrac{\partial f}{\partial t}+v_x\dfrac{\partial f}{\partial
x}=$$$$=-\dfrac{ie}{\hbar}W[U,f]+\nu \Big[f_{eq}(x,\mathbf{p},t)-
f(x,\mathbf{p},t)\Big],
\eqno{(1.3)}
$$

$$
W[U,f]=\dfrac{1}{(2\pi)^3}\int \Big[U\Big(x+\dfrac{\hbar
{b_x}}{2},t\Big)-
U\Big(x-\dfrac{\hbar {b_x}}{2},t\Big)\Big]
\times
$$
$$
\times
f(x,\mathbf{p'},t)\exp(i\mathbf{b}(\mathbf{p'}-\mathbf{p}))\,d^3
b\, d^3p'.
\eqno{(1.4)}
$$

We will carry out linearization of the equations (1.3) and (1.4).
Unperturbed  absolute Fermi --- Dirac distribution function
for degenerate plasma has the form
$$
f_F(p)=\Theta(\E_F-\E),
$$
where $\E_F$ is the kinetic energy of electron on
the Fermi  surface,
$$
\E_F(p)=\dfrac{mv_F^2}{2}=\dfrac{p_F^2}{2m}.
$$

In linear approximation in expression (1.4) instead of $f$ it is necessary
to take the absolute Fermi --- Dirac distribution function
$f_F$. Our linearization of the Wigner function for electrons and
the equilibrium distribution function leads to equalities:
$$
f=f_F(p)+U_0e^{i(kx-\omega t)}f_1(\mathbf{p}),
\eqno{(1.5)}
$$
$$
f_{eq}=f_F(p)+ \delta(\E_F-\E)\delta \E_F(x,t),
\eqno{(1.6)}
$$
where $f_1(\mathbf{p})$ is a new unknown function,
$\delta(x)$ is the Dirac delta--function,
$$
\delta \E_F(x,t)=\E_F(x,t)-\E_F,
$$
$U_0$ is the potential amplitude. We assume that has the form  of the
traveling wave
$$
U(x,t)=U_0e^{i(kx-\omega t)}.
\eqno{(1.7)}
$$

The quantity $\delta \E_F(x,t)$ describes
local change of Fermi's energy of the electronic gas, caused by change
of its density. Presence of this term in  collisions integral
provides realization of the particle number conservation law for
electrons.

Let's substitute (1.5) and (1.6) in the equation (1.3).
We receive the following equation:
$$
f_1(\mathbf{p})\Big[\nu-i\omega+ikv_x)\Big]U_0e^{i(kx-\omega t)}=$$$$=-
\dfrac{ie}{\hbar}W[U, f_F]+\nu \delta(\E_F-\E)
(\delta \E_F(x,t)).
\eqno{(1.8)}
$$

The Wigner --- Vlasov functional has the following
form in linear approximation:
$$
W[U,f_F]=\dfrac{1}{(2\pi)^3}\int \Big[U\Big(x+\dfrac{\hbar
{b_x}}{2},t\Big)-
U\Big(x-\dfrac{\hbar {b_x}}{2},t\Big)\Big]
\times
$$
$$
\times
f_F(p')\exp(i\mathbf{b}(\mathbf{p'}-\mathbf{p}))\,d^3
b\, d^3p',
\eqno{(1.9)}
$$

We derive following expression for potential:
$$
U(x+\dfrac{\hbar b_x}{2},t)-U(x-\dfrac{\hbar
b_x}{2},t)=$$$$=
U_0e^{i(kx-\omega t)}\Big[\exp(i\dfrac{k\hbar b_x}{2})-
\exp(-i \dfrac{k\hbar b_x}{2})\Big].
\eqno{(1.10)}
$$

We will integrate in (1.9) by $d^3b$.
Considering (1.10), we deduce:
$$
\dfrac{1}{(2\pi)^3}\int \Big[U\Big(x+\dfrac{\hbar b_x}{2}\Big)-
U\Big(x-\dfrac{\hbar b_x}{2}\Big)\Big]e^{i\mathbf{b}(\mathbf{p'}-
\mathbf{p})}d^3b=
$$
$$
=\dfrac{U(x,t)}{(2\pi)^3}\int \Big[\exp(i\dfrac{k\hbar b_x}{2})-
\exp(i \dfrac{k\hbar b_x}{2})\Big]e^{-i\mathbf{b}(\mathbf{p'}-\mathbf{p})}\,
d^3b=
$$
$$
=U(x,t)\delta(p_y'-p_y)\delta(p_z'-p_z) \times $$$$ \times
\Big[\delta\Big(p_x'-p_x+\dfrac{\hbar k}{2}\Big)-
\delta\Big(p_x'-p_x-\dfrac{\hbar k}{2}\Big)\Big]=
$$
$$
=U(x,t)\delta(p_y-p_y')\delta(p_z-p_z')\times $$$$ \times
\Big[\delta\Big(p_x-p_x'-\dfrac{\hbar k}{2}\Big)-
\delta\Big(p_x-p_x'+\dfrac{\hbar k}{2}\Big)\Big].
$$

It is necessary to integrate by momentums:
$$
W[U,f_F]=U(x,t)\int \delta(p_y-p_y')\delta(p_z-p_z') \times $$$$
\times
\Big[\delta\Big(p_x-p_x'-\dfrac{\hbar k}{2}\Big)-
\delta\Big(p_x-p_x'+\dfrac{\hbar k}{2}\Big)\Big] \times
$$
$$
\times \Theta\Big(\dfrac{p_F^2}{2m}-\dfrac{{p'}^2}{2m}\Big)\,dp_x'dp_y'dp_z'.
$$

As a result of integration by momentums we obtain:
$$
W[U,f_F]=U_0e^{i(kx-\omega t)} \times $$$$ \times
\Big[\Theta\Big(\dfrac{p_F^2}{2m}-\dfrac{p_y^2+p_z^2}{2m}-
\dfrac{\big(p_x-\frac{\hbar k}{2}\big)^2}{2m}\Big)-
$$
$$
-
\Theta\Big(\dfrac{p_F^2}{2m}-\dfrac{p_y^2+p_z^2}{2m}-
\dfrac{\big(p_x+\frac{\hbar k}{2}\big)^2}{2m}\Big)\Big],
$$
or
$$
W[U,f_F]=U_0e^{i(kx-\omega t)} \times $$$$ \times
\Big\{\Theta\Big[v_F^2-
\big(v_x-\frac{\hbar k}{2m}\big)^2-(v_y^2+v_z^2)\Big]-$$$$-
\Theta\Big[v_F^2-
\big(v_x+\frac{\hbar k}{2m}\big)^2-(v_y^2+v_z^2)\Big]\Big\}=
$$$$=U_0e^{i(kx-\omega t)}\Big[
\Theta_+(\mathbf{v})-\Theta_-(\mathbf{v})\Big],
$$
where
$$
\Theta_+(\mathbf{v})=\Theta\Big[v_F^2-
\Big(v_x-\frac{\hbar k}{2m}\Big)^2-(v_y^2+v_z^2)\Big],
$$
$$
\Theta_-(\mathbf{v})=\Theta\Big[v_F^2-
\Big(v_x+\frac{\hbar k}{2m}\Big)^2-(v_y^2+v_z^2)\Big],
$$

So, the Wigner --- Vlasov's functional is equal to:
$$
W[U,f_M]=U_0e^{i(kx-\omega t)}\Big[\Theta_+(\mathbf{v})-
\Theta_-(\mathbf{v})\Big].
\eqno{(1.11)}
$$

The quantity $ \delta \E_F (x, t)$ we will find from the conservation law
 of particle number:
$$
\int \nu(f_{eq}-f)d\Omega_F=0,
$$
where
$$
d\Omega_F=\dfrac{2\;d^3p}{(2\pi \hbar)^3}.
\eqno{(1.12)}
$$
According to the equality (1.12) we get the following equation:
$$
\int \nu(f_{eq}-f)d^3v=$$$$=\int \Big[\delta\E_F(x,t)
\delta(\E_F-\E)-U_0e^{i(kx-\omega t)}
f_1(\mathbf{p})\Big]\,d^3v=0.
$$

From this equality we obtain:
$$
\delta\E_F(x,t)=U_0e^{i(kx-\omega t)}\dfrac{\displaystyle\int
f_1(\mathbf{p})d^3v}{\displaystyle\int
\delta(\E_F-\E)d^3v}.
\eqno{(1.13)}
$$

The denominator of the expression (1.13) is equal to the following:
$$
\int \delta(\E_F-\E)d^3v=\dfrac{1}{mv_F}
\int \delta(v_F-v)d^3v=\dfrac{4\pi v_F}{m}.
$$

According to the equality (1.13) we have:
$$
\delta\E_F(x,t)=U_0e^{i(kx-\omega t)}\dfrac{m}{4\pi v_F}
\int f_1(\mathbf{p})d^3v.
\eqno{(1.14)}
$$

Substituting the expression (1.14) in the equation (1.8), we obtain:
$$
U_0e^{i(kx-\omega t)}f_1(\mathbf{p})\Big(\nu-i\omega+ikv_x)\Big)=
-\dfrac{ie}{\hbar}W[U,f_F]+
$$
$$
+U_0e^{i(kx-\omega t)}\dfrac{m \nu \delta(\E_F-\E)}
{4\pi v_F}\int f_1(\mathbf{p})d^3v.
\eqno{(1.15)}
$$

Let's rewrite the equation (1.15) with the help of (1.11) in a form:
$$
f_1(\mathbf{p})\Big(\nu-i\omega+ikv_x\Big)=-\dfrac{ie}{\hbar}
\Big[\Theta_+(\mathbf{v})-\Theta_-(\mathbf{v})\Big]+
$$
$$
+\dfrac{m \nu\delta(\E_F-\E)}
{4\pi v_F}\int f_1(\mathbf{p})d^3v.
\eqno{(1.16)}
$$

Taking into account the following equality
$$
\delta(\E_F-\E)=\dfrac{1}{mv_F}\delta(v_F-v)
$$
from the equation (1.16) we obtain:
$$
f_1(\mathbf{p})=-\dfrac{ie}{\hbar}\dfrac{\Theta_+(\mathbf{v})-
\Theta_-(\mathbf{v})}{\nu+i(k v_x-\omega)}+$$$$+
\dfrac{A\nu}{4\pi v_F^2}\dfrac{\delta(v_F-v)}{\nu+i(k v_x-\omega)}.
\eqno{(1.17)}
$$\\

\section{III. Longitudinal permeability and conductivity}

Let's designate:
$$
A=\int f_1(\mathbf{p})d^3v.
\eqno{(2.1)}
$$

Substituting (1.17) in the relationship (2.1), we get:
$$
A=-\dfrac{ie}{\hbar}\int \dfrac{\Theta_+(\mathbf{v})-
\Theta_-(\mathbf{v})}{\nu+i(k v_x-\omega)}d^3v+$$$$+
\dfrac{A\nu}{4\pi v_F^2}\int \dfrac{\delta(v_F-
v)d^3v}{\nu+i(k v_x-\omega)}.
\eqno{(2.2)}
$$

The last integral in (2.2) is easily calculated with the use of spherical
coordinates:
$$
\int\dfrac{\delta(v_F-v)d^3v}{\nu+i(k v_x-\omega)}=
\int\limits_{-1}^{1}\int\limits_{0}^{2\pi}\int\limits_{0}^{\infty}
\dfrac{v^2\delta(v_F-v)d\mu d\chi dv}{\nu+i(kv \mu-\omega)}=
$$
$$
=2\pi v_F^2
\int\limits_{-1}^{1}\dfrac{d\mu}{\nu+i(kv_F\mu-\omega)}=2\pi v_F^2
\dfrac{i}{kv_F}\ln\dfrac{\omega+i \nu+kv_F}{\omega+i \nu-kv_F}.
$$

Let's designate further:
$$
g_0(\omega,k,\nu)=\dfrac{i \nu}{2kv_F}\ln
\dfrac{\omega+i \nu+kv_F}{\omega+i \nu-kv_F}.
$$

Now from the equation (2.2) we obtain a relationship:
$$
A=-\dfrac{ie}{\hbar}\cdot\dfrac{1}{1- g_0(\omega,k,\nu)}
\int \dfrac{\Theta_+(\mathbf{v})-\Theta_-(\mathbf{v})}{\nu+i(k v_x-
\omega)}d^3v.
\eqno{(2.3)}
$$

Let's consider the integral from (2.3)
$$
J(\omega,k,\nu)=\int \dfrac{\Theta_+(\mathbf{v})-
\Theta_-(\mathbf{v})}{\nu+i(k v_x-
\omega)}d^3v.
$$

According to definition of Heaviside function we have:
$$
\Theta_{\pm}(\mathbf{v})=\left\{\begin{array}{c}
              1,\qquad \Big(v_x\mp \frac{\hbar k}{2m}\Big)^2+v_y^2+
              v_z^2 \leqslant v_F^2, \\
              0, \qquad \Big(v_x\mp \frac{\hbar k}{2m}\Big)^2+v_y^2+
              v_z^2 > v_F^2.
            \end{array}\right.
$$

Hence, this integral is equal to the following:
$$
J(\omega,k,\nu)=\int\limits_{S^3_+}\dfrac{ dv_x dv_y dv_z}
{\nu+i(k v_x-\omega)}-
\int\limits_{S^3_-}\dfrac{ dv_x dv_y dv_z}
{\nu+i(k v_x-\omega)}.
$$
or
$$
J(\omega,k,\nu)=J^+(\omega,k,\nu)-J^-(\omega,k,\nu),
$$
where
$$
J^{\pm}(\omega,k,\nu)=\int\limits_{S^3_{\pm}}\dfrac{ dv_x\,dv_y\,dv_z}
{\nu+i(k v_x-\omega)}.
$$

Here $S^3_{\pm}$ is a sphere with the centre in the point
$(\mp\frac{\hbar k}{2m},0,0)$.
The radius of this sphere is equal to electron velocity on Fermi's
surface,
$$
S^3_{\pm}= \left\{(v_x,v_y,v_z): \quad
\Big(v_x\pm \dfrac{\hbar k}{2m}\Big)^2+v_y^2+v_z^2\leqslant v_F^2\right\}.
$$

After obvious replacement of a variable
$v_x\pm \frac{\hbar k}{2m}\to v_x$
we receive for integrals $J^{\pm}$:

$$
J^{\pm}(\omega,k,\nu)= \int\limits_{S^3(0)}\dfrac{ dv_x\,dv_y\,dv_z}
{\nu+ik\big(v_x\pm \frac{\hbar k}{2m}\big)-i\omega},
$$
where  $S^3$ is the Fermi's sphere with the centre in
the beginning of coordinates,
$$
S^3(\mathbf{0})=S^3(0,0,0)= \left\{(v_x,v_y,v_z): \;
v_x^2+v_y^2+v_z^2\leqslant v_F^2\right\}.
$$

Fermi's sphere $S^3(0)$ we will present in the form:
$$
S^3(0)=\bigcup\limits_{v_x=-v_F}^{v_x=v_F}S^2_{v_F^2-v_x^2}(0,0).
$$

Here $S^2_{v_F^2-v_x^2}(0,0)$ there is a circle of the following form:
$$
S^2_{v_F^2-v_x^2}(0,0)= \left\{(v_y,v_z): \; v_y^2+v_z^2<v_F^2-v_x^2\right\}.
$$

Now  we will calculate integrals $J^{\pm}$ as repeated:
$$
J^{\pm}(\omega,k,\nu)=
\int\limits_{-v_F}^{v_F}
\dfrac{dv_x}{\nu+ik \big(v_x\pm \frac{\hbar
k}{2m}\big)-i\omega}\times $$$$ \times
\iint\limits_{S^2_{v_F^2-v_x^2}(0,0)}\;dv_y\,dv_z=
\pi\int\limits_{-v_F}^{v_F}
\dfrac{(v_F^2-v_x^2)dv_x}
{\nu+ik \big(v_x\pm \frac{\hbar k}{2m}\big)-i\omega}.
$$

Now these integrals can be calculated easily:
$$
J^{\pm}(\omega,k,\nu)=\dfrac{2i \pi \nu v_F}{k^2}(\omega_{\mp}+i \nu)-
$$$$-
\dfrac{i\pi \nu}{k^3}\Big[(\omega_{\mp}+i \nu)^2-k^2v_F^2\Big]
\ln \dfrac{\omega_{\mp}+i \nu +kv_F}
{\omega_{\mp}+i \nu -kv_F}.
$$

Here, as earlier,
$$
\omega_{\pm}=\omega\pm \dfrac{\hbar k^2}{2m}.
$$

The difference of integrals $J^+$ and $J^-$ is equal to:
$$
J(\omega,k,\nu)=-\dfrac{2i \pi \nu v_F \hbar}{m}+$$$$+\dfrac{i \pi \nu}{k^3}
\Big[(\omega_{+}+i \nu)^2-k^2v_F^2\Big]
\ln \dfrac{\omega_{+}+i \nu +kv_F}{\omega_{+}+i \nu -kv_F}
-
$$
$$
-\dfrac{i \pi \nu}{k^3}
\Big[(\omega_{-}+i \nu)^2-k^2v_F^2\Big]
\ln \dfrac{\omega_{-}+i \nu +kv_F}{\omega_{-}+i \nu-kv_F}.
\eqno{(2.4)}
$$

Let's present a difference (2.4) in the form:
$$
J(\omega,k,\nu)=$$$$=
-\dfrac{2i \pi \nu v_F \hbar}{m}\Big[1-g(\omega_+,k,\nu)+
g(\omega_-,k,\nu)\Big],
\eqno{(2.5)}
$$
where
$$
g(\omega_{\pm},k,\nu)=$$$$=
\dfrac{m\big[(\omega_{\pm}+i \nu)^2-k^2v_F^2\big]}{2\hbar k^3 v_F}
\ln \dfrac{\omega_{\pm}+i \nu +kv_F}{\omega_{\pm}+i \nu-kv_F}.
\eqno{(2.6)}
$$

Thus, the quantity $A$ according to (2.3) and (2.5) is equal to:
$$
A=-\dfrac{ie}{\hbar}\dfrac{J(\omega,k,\nu)}{1- g_0(\omega,k,\nu)}.
$$

Hence, according to  both  (1.17) and (2.5) function
$f_1(\mathbf {p})$ is constructed also:
$$
f_1(\mathbf{p})=-\dfrac{ie}{\hbar} \Bigg[
\dfrac{\Theta_+(\mathbf{v})-
\Theta_-(\mathbf{v})}{\nu+i(k v_x-\omega)}+\hspace{5cm}
$$
$$\qquad
+\dfrac{J(\omega,k,\nu)}{4\pi v_F^2(1- g_0(\omega,k,\nu))}
\cdot\dfrac{\delta(v_F-v)}
{\nu+i(kv_x-\omega)}\Bigg].
\eqno{(2.7)}
$$

Let's consider a relationship between electric field and potential
$$
\mathbf{E}(x,t)=-{\rm grad}\; U(x,t),
$$
or
$$
\mathbf{E}(x,t)=-\Big\{\frac{\partial U(x,t)}{\partial x},0,0\Big\},
$$
and the equation of a continuity for current and charge
densities:
$$
\dfrac{\partial \rho}{\partial t}+
\dfrac{\partial j_x}{\partial x}=0.
$$

Here according to definition
of dielectric conductivity we may represent the current density in the form:
$$
j_x= \sigma_l E_x=-\sigma_l \dfrac{\partial U}{\partial x}
=$$$$=-\sigma_l U_0ik e^{i(kx-\omega t)}=-\sigma_l ik U(x,t).
$$
Hence,
$$
\dfrac{\partial j_x}{\partial x}=\sigma_lk^2U(x,t).
$$

Taking into account  obvious equality for charge density
$$
\rho=e\int fd\Omega_F=
e\int [f_0(\E)+U_0e^{i(kx-\omega t)}
f_1]d\Omega_F,
$$
we obtain:
$$
\dfrac{\partial \rho}{\partial t}=-i\omega eU(x,t)\int
f_1 d\Omega_F.
$$

Substituting last two equalities in the continuity equation,
we find the general
formula for calculation of longitudinal conductivity:
$$
\sigma_l=\dfrac{ie\omega}{k^2}\int f_1d\Omega_F=
\dfrac{2ie\omega m^3}{(2\pi \hbar)^3 k^2}\int f_1 d^3v.
\eqno{(2.8)}
$$

Substituting (2.7) in (2.8), we get:
$$
\sigma_l=\dfrac{2e^2 \tau \omega m^3}{(2\pi \hbar)^3 k^2\hbar}
\Bigg[\int
\dfrac{(\Theta_+(\mathbf{v})-\Theta_-(\mathbf{v}))\,d^3v}
{\nu+i(k v_x-\omega)}+\qquad
$$$$\qquad+
\dfrac{J(\omega,k,\nu)}{4\pi v_F^2(1-g_0(\omega,k,\nu))}
\int \dfrac{\delta(v_F-v)d^3v}
{\nu+i(k v_x-\omega)}\Bigg].
$$
and, using formulas
$$
\int \dfrac{(\Theta^+(\mathbf{v})-\Theta^-(\mathbf{v}))d^3v}
{\nu+ik v_x-i\omega}=J(\omega,k,\nu),
$$
$$
\int \dfrac{\delta(v_F-v)d^3v}{\nu+ik v_x-i\omega}=4\pi
v_F^2 g_0(\omega,k,\nu),
$$
we derive:
$$
\sigma_l=\dfrac{i\pi e^2 \omega m^3 v_F}
{2\pi^3 \hbar^3\ m  k^2}
\cdot \dfrac{J(\omega,k,\nu)}{1-g_0(\omega,k,\nu)}.
$$

With the help of the formula for numerical electron
density of degenerate plasma
$$
\Big(\dfrac{mv_F}{\hbar}\Big)^3=3\pi^2N,
$$
following expression for calculation of longitudinal conductivity is
obtained:
$$
\sigma_l=-\dfrac{3i e^2 N\omega }{2m k^2 v_F^2}
\cdot \dfrac{1-g(\omega_+,k,\nu)+g(\omega_-,k,\nu)}
{1-g_0(\omega,k,\nu)},
\eqno{(2.9)}
$$
or, with the use of classical conductivity
$\sigma_0=\dfrac{e^2N}{m\nu}$,
this formula will be written in the form:
$$
\sigma_l=\sigma_0\cdot\Big(-\dfrac{3i}{2}\Big)
\cdot \dfrac{\omega \nu}{(kv_F)^2}
\cdot $$$$\cdot \dfrac{1-g(\omega_+,k,\nu)+g(\omega_-,k,\nu)}
{1-g_0(\omega,k,\nu)}.
\eqno{(2.9')}
$$

Using definition of dielectric permeability
$$
\varepsilon_l=1 +\dfrac {4\pi i} {\omega} \sigma_l,
$$
with the help of (2.9) we will get the following representation for
longitudinal dielectric permeability of plasma:
$$
\varepsilon_l(\omega,k,\nu)=$$$$=1+\dfrac{3\omega_p^2}{2k^2v_F^2}
\cdot \dfrac{1-g(\omega_+,k,\nu)+g(\omega_-,k,\nu)}
{1-g_0(\omega,k,\nu)},
\eqno{(2.10)}
$$
where $\omega_p$ is the electron plasma frequency,
$$
\omega_p=\dfrac{4\pi e^2 N}{m}.
$$

Let's enter dimensionless parameters:
$$
z=x+i y,\quad x=\dfrac{\omega}{kv_F},\quad y=\dfrac{\nu}{kv_F},
$$$$
x_p=\dfrac{\omega_p}{kv_F},\quad q=\dfrac{k}{k_F},
$$
where $k_F=\dfrac{mv_F}{\hbar}$  is the Fermi wave number.

The dimensionless dielectric function of these parameters has the
the following form:
$$
\varepsilon_l(x,y,q)=1+\dfrac{3}{2}x_p^2\dfrac{1-g(z,+q)+g(z,-q)}
{1-g_0(x,y)}.
\eqno{(2.11)}
$$

Here
$$
g_0(x,y)=\dfrac{iy}{2}\ln\dfrac{x+i y+1}{x+iy-1}, \qquad z=x+iy,
$$
$$
g(z,\pm q)=\dfrac{(z\pm q/2)^2-1}{2q}\ln\dfrac{z\pm q/2+1}{z\pm q/2-1}.
$$

Let $ \nu=0$, i.e. plasma is collisionless; then from the expression
(2.10) the following classical formula for the dielectric
permeability follows:
$$
\varepsilon_l(\omega,k)=1+\dfrac{3\omega_p^2}{2k^2v_F}
\Big[1-g(\omega_-,k)+g(\omega_+,k)\Big],
\eqno{(2.12)}
$$
where
$$
g(\omega_{\pm},k)=\dfrac{m(\omega_{\pm}^2-k^2v_F^2)}{2\hbar k^3v_F}\ln
\dfrac{\omega_{\pm}+kv_F}{\omega_{\pm}-kv_F}.
$$

This formula is called  (see, for example,
\cite{Kohn1} - \cite{Harr})
dielectric Lindhard's function \cite{Lind} in the literature.
It is deduced by the method of random phases approximation.

In dimensionless variables dielectric Lindhard's function has the
following form:
$$
\varepsilon_l(x,q)=1+\dfrac{3}{2}x_p^2[1-g(x,+q)+g(x,-q)],
\eqno{(2.13)}
$$
where
$$
g(x,\pm q)=\dfrac{(x\pm q/2)^2-1}{2q}\ln\dfrac{x\pm q/2+1}{x \pm q/2-1}.
$$

\begin{figure}[h]
\begin{center}
\includegraphics[width=8.5cm, height=4.5cm]{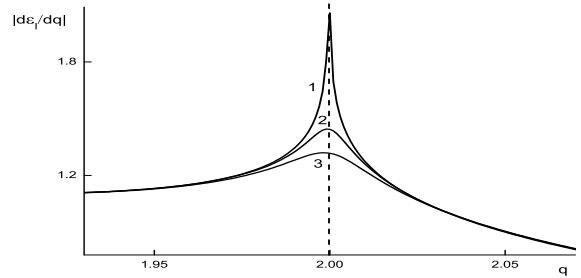}
\caption{Kohn's singularity, $x_p=1, \; x=0$; curves of $1,2,3$
correspond
to parameters which are values of dimensionless collision
frequencies
$y=0, \; 0.005 \; 0.01$.
}\label{rateI}
\end{center}
\end{figure}

On Fig. 1  Kohn's singularity in a case,
when $x_p=1, \; x=0$ is represented; curves of $1,2,3$
correspond
to parameters which are values of dimensionless collision
frequencies
$y=0, \; 0.005 \; 0.01$.

\begin{figure}[h]
\begin{center}
\includegraphics[width=8.2cm, height=4.5cm]{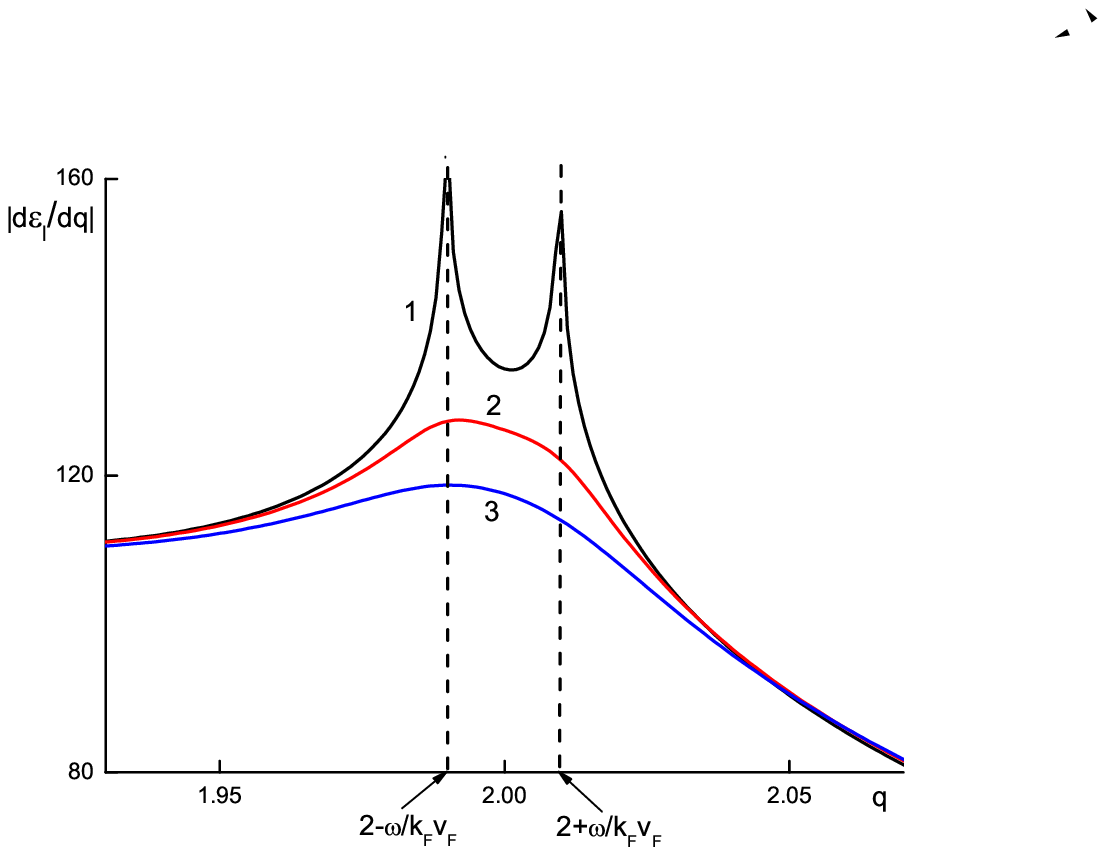}
\caption{Kohn's singularities, $x_p=10, \; x=0$,
curves of $1,2,3$ correspond to the
values of parameter $y=0; \; 0.01; \; 0.02$.
}\label{rateII}
\end{center}
\end{figure}

\begin{figure}[h]
\begin{center}
\includegraphics[width=8.0cm, height=4.5cm]{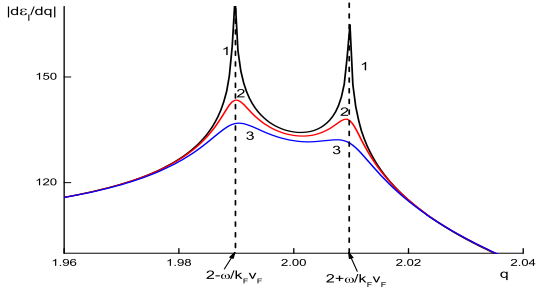}
\caption{Kohn's singularities, $x_p=10, \; x=0$,
curves of $1,2,3$ correspond to values of the parameter
$y=0; \; 0.002; \; 0.004$.
}\label{rateIII}
\end{center}
\end{figure}
\

On Fig. 2  Kohn's singularities in a case, when
$x_p=10, \; x=0$  are represented,  curves of $1,2,3$ correspond to the
values of parameter
$y=0; \; 0.01; \; 0.02$. On Fig. 3 the case, when
$x_p=10, \; x=0$ is represented, curves of $1,2,3$ correspond to
values of the parameter
$y=0; \; 0.002; \; 0.004$.

In the formula (2.11) we will write  new dimensionless variables:
$$
z=\dfrac{\omega+i \nu}{k_Fv_F}=x+iy, \qquad x=\dfrac{\omega}{k_Fv_F},
$$
$$
 y=\dfrac{\nu}{k_Fv_F}, \qquad q=\dfrac{k}{k_F}.
$$

At such replacement of variab\-les we obtain the following formula
for the longitudinal dielectric permeability:
$$
\varepsilon_l(x,y,q)=1+\dfrac{3x_p^2}{2q^2}\dfrac{1-g_+(z,q)+g_-(z,q)}
{1-g_0(x,y,q)},
\eqno{(2.13)}
$$
where
$$
x_p^2=\dfrac{\omega_p^2}{k_Fv_F},
$$ \medskip
$$
g_0(x,y,q)=\dfrac{iy}{2q}\ln\dfrac{x+iy+q}{x+iy-q},
$$\medskip
$$
g_+(z,q)=\dfrac{(z+q^2/2)^2-q^2}{2q^3}\ln\dfrac{z+q+q^2/2}
{z-q+q^2/2},
$$\medskip
$$
g_-(z,q)=\dfrac{(z-q^2/2)^2-q^2}{2q^3}\ln\dfrac{z+q-q^2/2}
{z-q-q^2/2}.
$$

Kohn's singularities are determined from four equations:
$$
q^2\pm 2q\pm2z=0.
\eqno{(2.14)}
$$

These equations at $y=0;(\nu=0)$ define four Kohn's singularities,
two of which at $\omega\ne 0$ lay in neighbourhood of point $q=2$:
$$
q_{1,2}=1\pm \sqrt{1\pm 2x},
$$
and two others lay in point neighbourhood $q =-2$:
$$
q_{3,4}=-1-\sqrt{1\pm 2x}.
$$

In the case of infinitesimal $x$ we have from these formulas:
$$
q_{1,2}\approx 2\pm x\approx 2\pm \dfrac{\omega}{k_Fv_F}, \qquad
q_{3,4}\approx -2\pm x\approx -2\pm \dfrac{\omega}{k_Fv_F}.
$$

In terms of dimensional Fermi wave number the Kohn's singularities
are determined by equalities:
$$
k_{1,2}=k_F+ \sqrt{k_F^2\pm 2\dfrac{k_F\omega}{v_F}},
$$
and
$$
k_{3,4}=-k_F-\sqrt{k_F^2\pm2\dfrac{k_F\omega}{v_F}}.
$$

Besides that, these formulas may  be rewritten in a form:
$$
k_{1,2}=
\dfrac{mv_F}{\hbar}\Big(1+\sqrt{1\pm 2\dfrac{\hbar
\omega}{mv_F^2}}\Big)=\dfrac{p_F}{\hbar}\Big(1+\sqrt{1\pm \dfrac{\hbar
\omega}{\E_F}}\Big)
$$
and
$$
k_{3,4}=\dfrac{mv_F}{\hbar}\Big(-1-\sqrt{1\pm 2\dfrac{\hbar
\omega}{mv_F^2}}\Big)=\dfrac{p_F}{\hbar}\Big(1+\sqrt{1\pm
\dfrac{\hbar \omega}{\E_F}}\Big).
$$

Here $\E_F$ is the electron kinetic energy on a Fermi's surface
$$
\E_F=\dfrac{mv_F^2}{2}.
$$

Thus, into collisionless plasma ($\nu=0$) at $\omega\ne 0$
there is a splitting of Kohn's singularities.

\section{III. Comparison with Mermin's result}

Mermin (see Mermin N.D. \cite {Mermin}) has obtaned
the following expression of dielectric function:
$$
\varepsilon^M(\omega,k)=
1+\dfrac{(\omega+i \nu)\Big[\varepsilon^\circ
(\omega+i \nu,k)-1\Big]}
{\omega +i\nu \dfrac{\varepsilon^\circ(\omega+i \nu,k)-1}
{\varepsilon^\circ(0,k)-1}}.
\eqno{(3.1)}
$$

The formula (3.1) is obtained on the basis of the kinetic equation
for one--partial density matrix  $ \rho $ in momentum space.

In the formula (3.1) the function
$\varepsilon^\circ(\omega,k)$ is the so-called
Lindhard's dielectric function, i.e.
the dielectric function obtained for collisionless plasma,
and defined by the equality (2.12):
$$
\varepsilon^\circ(\omega,k)\equiv
\varepsilon_l(\omega,k)=1+\dfrac{3\omega_p^2}{2k^2v_F}
\Big[1-g(\omega_+,k)+g(\omega_-,k)\Big].
$$

Expression $\varepsilon^\circ(\omega+i\nu,k)$
means that arguments of dielectric Lindhard function $\omega_{\pm}$
are replaced formally on $\omega_{\pm} + i \nu$, i.e.
$$
\varepsilon_l^\circ(\omega+i \nu,k)-1=$$$$=\dfrac{3\omega_p^2}{2k^2v_F^2}
\Big[1-g(\omega_++i \nu,k)+g(\omega_-+i \nu,k)\Big].
\eqno{(3.2)}
$$

So the function $\varepsilon^\circ(0,k)$ has the form
$$
\varepsilon_l^\circ(0,k)-1=\dfrac{3\omega_p^2}{2k^2v_F^2}
\Big[1-g(0_+,k)+g(0_-,k)\Big].
\eqno{(3.3)}
$$

Here
$$
g(0_{\pm},k)=\dfrac{m\Big(\frac{\hbar k^4}{4m^2}-k^2v_F^2\Big)}
{2\hbar k^3v_F}
\ln\dfrac{\pm\frac{\hbar k^2}{2m}+kv_F}
{\pm\frac{\hbar k^2}{2m}-kv_F}=
$$
$$
=\dfrac{m\Big(\frac{\hbar k^2}{4m^2}-v_F^2\Big)}
{2\hbar kv_F}
\ln\dfrac{\pm\frac{\hbar k}{2m}+v_F}
{\pm\frac{\hbar k}{2m}-v_F}.
$$

From these formulas follows
that these two functions differ only in sign:
$$
g(0_-,k)=-g(0_+,k).
$$

From formulas (3.2) and (3.3) one can see that
$$
\dfrac{\varepsilon^\circ(\omega+i
\nu,k)-1}{\varepsilon^\circ(0,k)-1}=$$$$=
\dfrac{1-g(\omega_++i \nu)+g(\omega_-+i \nu)}
{1-g(0_+,k)+g(0_-,k)}.
\eqno{(3.4)}
$$

With the help of (3.2)--(3.4) Mermin formula (3.1)
will be written in our designations in the following form:
$$
\varepsilon^M=1+\dfrac{3\omega_p^2}{2k^2v_F^2}\times
$$
$$
\dfrac{(\omega + i \nu)
[1-g(\omega_++i \nu,k)+g(\omega_-+i \nu,k)]}
{\omega+i \nu \dfrac{1-g(\omega_++i \nu,k)+g(\omega_-+i \nu,k)}
{1-g(0_+,k)+g(0_-,k)} }.
\eqno{(3.5)}
$$

From formulas (2.10) and (3.5) one can see, that in the case $\nu\to 0$
the formula deduced in this work and
Mermin formula turn into  the same expression for dielectric function
for quantum collisionless plasma that is Lindhard dielectric
function (2.12).

\section{IV. Conclusion}

It is interesting to notice, that in the case of low-frequency limit,
i.e. at $\omega=0$ Mermin dielectric  function does
not depend on collision frequency  $\nu $. Indeed, assuming $\omega=0$
in the formula (3.1), we obtain
$$
\varepsilon_l^M(0,k,\nu)=\varepsilon^\circ(0,k)=$$$$=
1+\dfrac{3\omega_p^2}{2k^2v_F^2}\Big[1-g(0_+,k)+g(0_-,k)\Big],
\eqno{(4.1)}
$$
or, in dimensionless variables,
$$
\varepsilon_l^M(0,k,\nu)=1+\dfrac{3}{2}x_p^2\Big[1-\dfrac{w^2-1}{2w}
\ln\dfrac{w+1}{w-1}\Big].
\eqno{(4.2)}
$$

From the formula obtained in the this work (2.10)
we have another formula in a low--frequency limit:

$$
\varepsilon_l(0,y,w)=1+\dfrac{3x_p^2}{2}\Big[1-\dfrac{iy}{2}
\ln\dfrac{iy+1}{iy-1}\Big]^{-1}\times
$$
$$
\times\Big[1-\dfrac{(iy+w)^2-1}{4w}\ln\dfrac{iy+w+1}{iy+w-1}+$$$$+
\dfrac{(iy-w)^2-1}{4w}\ln\dfrac{iy-w+1}{iy-w-1}\Big].
\eqno{(4.3)}
$$\\

The formula (4.3) transforms into the formula (4.2) at $y=0$.

So, in the present work analytical expression for the longitudinal
quantum dielectric permeability of degenerate electron plasma
is derived.
Kinetic  Wigner --- Vlasov ---  Boltzmann equation
with collision integral  in the form of relaxation
$\tau $ -- model in coordinate space  is used.

It is shown, that in a limiting case, when
Planck constant tends to zero, the expression
obtained is transformed in the classical formula for the
longitudinal dielectric permeability of degenerate plasma.

Static limits ($\omega\to 0$)
for the dielectric permeability for collisionless,
and for collisional plasma have been found.

Splitting of Kohn singularities in collisionless plasma is
marked.

Comparison with classical Mermin's result for
dielectric permea\-bi\-li\-ty has been carried out. We will notice,
that Mermin formula was obtained
with use of the relaxation kinetic equation in the
momentum space. For collisionless plasma the formula deduced in
this work, and the Mermin formula as well can be transformed into
the same Lindhard formula.

\end{document}